\documentclass{aa}
\def\gr{$\gamma$-ray}
\usepackage{natbib} 
\usepackage{graphicx}
\usepackage{color}
\usepackage{subcaption} 
\usepackage{tablefootnote}
\usepackage{multicol}

\begin{document}

\title{Revision of conservative lower bound on the intergalactic magnetic field from Fermi and Cherenkov telescope observations of extreme blazars}
\author{Jeffrey Blunier$^{1}$, Andrii Neronov$^{1,2}$, Dmitri Semikoz$^{1}$ }

\institute{
Universit\'e Paris Cit\'e, CNRS, Astroparticule et Cosmologie, 75006 Paris, France
\and
Laboratory of Astrophysics, \'Ecole Polytechnique F\'ed\'erale de Lausanne, 1015 Lausanne, Switzerland
}

\authorrunning{Blunier, Neronov \& Semikoz}
\titlerunning{Lower bound on IGMF from extreme blazars}

 \abstract
     {Joint observations of extreme blazars with the Fermi Large Area Telescope (LAT) and Imaging Atmospheric Cherenkov telescopes (IACTs) have previously been used to derive lower bounds on the intergalactic magnetic field (IGMF).}{We aim to update these previous bounds using a set of extreme blazars that were detected in the very-high-energy (VHE, photon energies above 100~GeV) band by both Fermi/LAT and IACTs. }{We measured IGMF-dependent suppression of secondary  delayed \gr\ flux from electron-positron pairs deposited in the intergalactic medium by VHE \gr s interacting with extragalactic background light. }{From a total of 22 extreme blazars detected by Fermi/LAT and IACTs in the VHE band,  seven have spectral characteristics inconsistent with the possibility of zero magnetic field along their lines of sight, even under the most restrictive assumption that the sources only switched on at the start of VHE band observations. Adopting this assumption, we derive a "conservative" lower bound on the IGMF strength at the level of $\sim 2\times 10^{-17}$~G. The tightest bound is imposed by the signal of 1ES 0502+675, a source that has not been considered in the IGMF analysis before. Our bound is comparable to the bound derived by MAGIC collaboration, but it is weaker than that previously derived from analyses of Fermi/LAT and HESS telescope data, even though our dataset includes those data. We clarify the origin of this discrepancy.  }{}

\maketitle

\section{Introduction}
\label{sec:intro}

Very-high-energy (VHE) \gr s from distant blazars (a type of active galactic nucleus (AGN) with jets aligned along the line of sight) interact with optical-infrared extragalactic-background-light (EBL) photons on their way from the source to the Earth and deposit electron-positron pairs in the intergalactic medium, mostly in the voids of the large-scale structure. The electrons and positrons subsequently lose energy via inverse Compton scattering of cosmic-microwave-background photons and EBL and produce secondary \gr\ emission detectable by \gr\ telescopes \citep{1994ApJ...423L...5A}.  In the absence of magnetic fields in the voids, the secondary \gr\ flux would be emitted in the direction of the primary \gr\ beam and provide an additional contribution to the blazar \gr\ flux. The intergalactic magnetic field (IGMF) present in the voids deflects the trajectories of electrons and positrons so that the secondary emission is no longer directed along the primary \gr\ beam, suppressing the secondary flux in an IGMF-dependent way \citep{1995Natur.374..430P,2007JETPL..85..473N}. This suppression has been observed \citep{2010Sci...328...73N} and used to establish the existence of nonzero magnetic field in the voids \citep{2010Sci...328...73N,2011A&A...529A.144T,2023A&A...670A.145A,2018ApJS..237...32A,HESS:2023zwb}.

\cite{2023A&A...670A.145A} derived a conservative lower bound $B>1.8\times 10^{-17}$~G on the void IGMF based on the non-observation of delayed secondary signal in the \gr\ flux of the blazar 1ES 0229+200, using a combination of data of MAGIC, HESS, and VERITAS Imaging Atmospheric Cherenkov Telescopes (IACTs) and the Fermi Large Area Telescope (LAT). The conservative assumption of  \cite{2023A&A...670A.145A} was that this source was never active in the VHE \gr\ band before the start of historical observations with IACTs two decades ago. Even though this is most probably not the case (the activity timescales, $T,$ of blazars span tens to hundreds of millions of years, as indicated by their large-scale jets), it cannot formally be ruled out that a specific source had just been "switched on" at the moment of birth of VHE astronomy. 

\cite{HESS:2023zwb} considered a combination of Fermi/LAT data with that of HESS IACT for five blazars and derived lower bounds on the IGMF, gradually relaxing the assumptions about the activity cycles of those sources: decade-long $T\sim 10$~yr (similar to \cite{2023A&A...670A.145A}), $T\sim 10^4$~yr, or $T\sim 10^7$~yr. The bounds derived by \cite{HESS:2023zwb} are much stronger, in particular for the "conservative" timescale of $T\sim 10$~yr, $B>7\times 10^{16}$~G. In the specific case of 1ES 0229+200, the lower bound derived by \cite{HESS:2023zwb} is $B>4\times 10^{-16}$~G, which is a factor of twenty better than that of \cite{2023A&A...670A.145A}. This is surprising, given that the analysis of \cite{2023A&A...670A.145A} also includes most of the HESS data considered by \cite{HESS:2023zwb}. 

The flux of the secondary \gr\ emission from the intergalactic medium is determined by the level of the primary VHE source flux in multi-TeV energy range. This flux is maximized for the sources with hard-VHE \gr\ spectra and high-VHE flux extending without high-energy cut-off in the TeV range, provided that the $\gamma\gamma$ opacity of the EBL is sufficient to convert primary photons into secondary pairs. The latter may result in a negligible cascade component for nearby sources despite having a hard primary spectrum. Sources exhibiting these spectral characteristics are known as "extreme" blazars \citep{1998MNRAS.299..433F}. The VHE emission from the extreme blazars is believed to be produced through a synchrotron-self-Compton mechanism via the up-scattering of the synchrotron photons by energetic electrons. The frequency of the high-energy cut-off in the synchrotron spectrum correlates with the energy of the cut-off in the \gr\ spectrum. The extreme blazars are those with the highest energy synchrotron cut-off frequencies (in excess of $10^{17}$~Hz) \citep{2019A&A...632A..77C}. The extreme blazars detected in the VHE band, the "extreme TeV blazars" \citep{2020NatAs...4..124B}, are  the best targets for the IGMF analysis. 

\citet{2025arXiv250608497N} compiled a catalog of VHE-emitting AGNs based on a long exposure of Fermi/LAT. This catalog lists 275 such AGNs in the sky region at Galactic latitudes $|b|>10^\circ$. The catalog includes 63 extreme blazars. 22 of them have already been detected by IACTs.

In what follows, we consider the IACT-detected extreme blazars  to reevaluate the "conservative" lower bound on the IGMF. In parallel, we were able to clarify the origin of discrepancy between the bound derived by \cite{2023A&A...670A.145A} and \cite{HESS:2023zwb}. Our analysis is an improvement on both works, because it benefits from better knowledge of   the "best targets" for the IGMF analysis, which were not considered in those two works. 

\section{Very-high-energy-detected extreme-blazar set}

Table \ref{tab:sources} presents the list of extreme blazars detected in the VHE band by both Fermi/LAT and IACTs \citep{2025arXiv250608497N}.  As mentioned in the introduction, there are a further 41 VHE-detected extreme blazars reported by \citet{2025arXiv250608497N}, for which there are currently no IACT data available. For these sources we cannot yet judge if they provide promising targets for the IGMF analysis, and we did not consider them in this work. We also did not consider the brightest source, Mrk 501, for which dedicated studies have found that the secondary flux is  largely subdominant \citep{Takahashi:2011ac,2021ApJ...906..116K}. 

We combined broad-band GeV-TeV spectra of the 21 sources from Table \ref{tab:sources} by complementing Fermi/LAT measurements with measurement by IACTs reported in the literature (cited in the "Time" column of the table). To extract Fermi/LAT spectra we used the aperture photometry approach in the same way as \citet{2025arXiv250608497N} (see Appendix \ref{sec:comparison} for details). We fit all the combined Fermi/LAT $+$ IACT spectra with cut-off power-law model, $N (E/1\mbox{ TeV})^{-\Gamma}\exp(-E/E_{cut}),$ which was modified by the effect of absorption on EBL and taken from \cite{2021MNRAS.507.5144S}. We derived values of the slope, $\Gamma$; cut-off energy, $E_{cut}$; and normalization of the spectrum, $N,$ for each source. The values of $\Gamma,E_{cut}$ are given in  columns 5 and 6 of  Table \ref{tab:sources}. For most of the extreme blazars, the spectrum extends well into the TeV energy range, creating a degeneracy in the fit for the cut-off energy after a certain threshold. Consequently, we only considered the lowest cut-off energy compatible with the fit and defined it as a lower bound, as can be seen from the "$E_{cut}$" column of the table.

\begin{table*}
\caption{List of sources used in this work.}
\begin{tabular}{lllllllllllll}
\hline
&Name               & RA & Dec & Time  & $\Gamma$  & $E_{cut}$ &$z$ &$\chi_\infty^2/dof$ &$\chi_0^2/dof$\\ 
\hline
1  &  SHBL J001355.9-185406  &  3.48  &  -18.912  & 41.5h H\tablefootmark{1}  &  $1.68_{-0.12}^{+0.08}$  & $0.8_{-0.3}^{+0.5}$ & 0.095  &2.86/10 & 3.3/10\\
2  &  1ES 0229+200  &  38.214  &  20.316  &  144.1h H\tablefootmark{2}   &  $1.63_{-0.05}^{+0.03}$ & $>3.6$ &  0.14  & $13.9/16$ & $50.1/16$ \\
  &    &    &    &  54h V\tablefootmark{3}  &  & &      \\
 &    &    &    &  145.5h M\tablefootmark{4}  &  & &      \\
3  &  RBS 0413  &  49.972  &  18.753  &  48h V\tablefootmark{5} &  $1.72_{-0.17}^{+0.15}$  &  $>0.8$ &0.19 & 8.4/11 & 6.3/11 \\
4  &  1ES 0347-121  &  57.354  &  -11.994  & 59.2h H\tablefootmark{2}  &  $1.62_{-0.06}^{+0.05}$  & $>0.9$ & 0.188  &29.1/16 & 37.1/16   \\
5  &  1ES 0502+675  &  76.996  &  67.622  & 13h V\tablefootmark{6}  & $1.50_{-0.06}^{+0.03}$  & $>1.0$ & 0.314   & 17.7/14 & 85.3/14 \\
6  &  PKS 0548-322  &  87.625  &  -32.277  & 53.9h H\tablefootmark{2} &  $1.80_{-0.03}^{+0.03}$ &  $>3.7$ &0.069 & 26.3/16 & 24.7/16\\
7  &  RGB J0710+591  &  107.623  &  59.135  &  22.1h V\tablefootmark{7}   &  $1.76_{-0.05}^{+0.05}$  & $>3.2$& 0.125  &26.3/11 & 26.8/11 \\
8  &  PGC 2402248  &  113.362  &  51.88  & 50h M\tablefootmark{8}  &  $1.78_{-0.02}^{+0.03}$  & $>1.3$ & 0.065  & 9.2/11 & 9.7/11  \\
9  &  RBS 0723  &  131.812  &  11.569  & 45.3h M\tablefootmark{9} &  $1.57_{-0.15}^{+0.12}$ & $>0.2$  &  0.198  & 6.2/8 & 5.2/8  \\
10  &  MRC 0910-208  &  138.227  &  -21.045  &  17h H\tablefootmark{10}  &  $1.75_{-0.08}^{+0.05}$  & $0.6_{-0.2}^{+0.5}$ & 0.198 & 5.6/9 & 5.0/9  \\
11  &  1ES 1101-232  &  165.909  &  -23.496  &  71.9h H\tablefootmark{2} &  $1.58_{-0.04}^{+0.04}$ & $>2.0$  &  0.186  & 10.3/15 & 44.2/15  \\
12  &  RX J1136.5+6737  &  174.118  &  67.613  &  30h M\tablefootmark{11}  &  $1.86_{-0.06}^{+0.06}$ &$>1.0$ &  0.134  & 7.9/10 & 7.7/10   \\
13  &  1ES 1312-423  &  198.768  &  -42.611  & 150h H\tablefootmark{12} &  $1.69_{-0.12}^{+0.09}$  & $0.8_{-0.4}^{+0.6}$ & 0.105    & 23.6/12 & 25.6/12   \\
14  &  RBS 1366  &  214.494  &  25.724  &  56h V\tablefootmark{13}   &  $1.71_{-0.09}^{+0.05}$  &  $>0.4$ & 0.236 & 12.7/12 & 14.7/12  \\
15  &  H 1426+428  &  217.129  &  42.678  & 73.5h V\tablefootmark{14}   &  $1.795_{-0.015}^{+0.015}$ & $>3.2$ &  0.129  & 22.1/17 & 26.9/17   \\
16  &  1ES 1440+122  &  220.698  &  12.013  &  53h V\tablefootmark{15}   &  $1.77_{-0.08}^{+0.09}$  & $>0.3$ &0.163 & 5.8/11 & 6.6/11  \\
17  &  1ES 1727+502  &  262.078  &  50.227  &  508d L\tablefootmark{16} &  $1.83_{-0.06}^{+0.08}$ & $2.0_{-0.4}^{+1.3}$ &  0.055   & 17.2/12 & 20.0/12  \\
18  &  1ES 1741+196  &  266.008  &  19.596  &  30h V\tablefootmark{17} &  $1.96_{-0.06}^{+0.06}$  & $>0.4$ & 0.08  & 7.0/10 & 6.9/10  \\
19  &  1RXS J195815.6-301119  &  299.581  &  -30.181  & 7.3h H\tablefootmark{10} &  $1.90_{-0.05}^{+0.03}$ & $>0.9$ &  0.119 & 5.5/11 & 5.7/11  \\
20  &  MG2 J204208+2426  &  310.536  &  24.458  &   52.5h M\tablefootmark{9}  &  $1.95_{-0.05}^{+0.05}$  & $>0.6$ &  0.104  & 14.1/9 & 13.7/9  \\
21  &  H 2356-309  &  359.772  &  -30.637  & 150.5h H\tablefootmark{2}  &  $1.63_{-0.05}^{+0.03}$ &$1.5_{-0.5}^{+1.3}$ &  0.165 & 20.4/16  & 26.1/16  \\
\hline
\end{tabular}
\tablefoot{The third and fourth columns give source coordinates. The fifth column provides information on the already-available exposures of different IACTS ("H" for HESS, "M" for MAGIC, and "V" for VERITAS). The only exception is 1ES 1727+502, for which we mention LHAASO detection (letter "L" is for LHAASO). The sixth and seventh columns provide spectral parameters (spectral slope, $\Gamma,$ and cut-off energy, $E_{cut}$, in TeV). The last column provides the source redshift.\\
\tablefoottext{1}{\citet{2013AA...554A..72H};}
\tablefoottext{2}{\citet{HESS:2023zwb};}
\tablefoottext{3}{\citet{2013ICRC...33.1105M};}
\tablefoottext{4}{\citet{2023A&A...670A.145A};}
\tablefoottext{5}{\citet{2012ApJ...750...94A};}
\tablefoottext{6}{\citet{Orr:2013yma};}
\tablefoottext{7}{\citet{2010ApJ...715L..49A};}
\tablefoottext{8}{\citet{2019MNRAS.490.2284M};}
\tablefoottext{9}{\citet{2020ApJS..247...16A};}
\tablefoottext{10}{\citet{2022icrc.confE.823B};}
\tablefoottext{11}{\citet{Hayashida:2015erl};}
\tablefoottext{12}{\citet{2013MNRAS.434.1889H};}
\tablefoottext{13}{\citet{Ribeiro:2023jgj};}
\tablefoottext{14}{\citet{VERITAS_H_1426+428};}
\tablefoottext{15}{\cite{2016MNRAS.461..202A};}
\tablefoottext{16}{\citet{2024ApJS..271...25C};}
\tablefoottext{17}{\cite{2016MNRAS.459.2550A}.}
} 
\label{tab:sources}
\end{table*}

\section{Secondary emission from the intergalactic medium}

The cut-off power-law fit to the source spectra corresponds to the physical model with an "infinite" IGMF; i.e., with an IGMF strong enough to completely suppress the secondary flux from the direction of the source. Another extreme is the case with zero IGMF. For each source, we attempted to fit such zero-IGMF alternative models to the spectrum. In this model, the total flux is a sum of the primary source flux attenuated by the effect of pair production on EBL plus the secondary flux from the inverse Compton scattering by the electron positron pairs. In the absence of an IGMF, the secondary flux is emitted in the same direction as the primary source flux and contributes to the point-source spectrum.  There is also (almost) no time delay of the secondary signal, so even if all the sources from Table \ref{tab:sources} only appeared on the sky 20 years ago (with the start of observations by IACTs), their signals should be accompanied by the secondary emission component. In a more general situation, the secondary \gr s arrive with a delay \citep{2009PhRvD..80l3012N},
\begin{equation}
T_d=2\times 10^2(1+z)^{-5}\left[\frac{E_\gamma}{100\mbox{ GeV}}\right]^{-5/2}
\left[\frac{B}{10^{-16}\mbox{ G}}\right]^2\mbox{ yr}
\label{eq:delay}
,\end{equation}
in the case of magnetic field with a correlation length of $\lambda_B\gg l_e$, where $l_e$ is the cooling distance of electrons and positrons with respect to the inverse Compton scattering.  Considering $T\sim 16$~yr for our (over)conservative assumption about the source-activity timescale (the time span of Fermi/LAT data considered in our analysis), one arrives at an order-of-magnitude estimate of suppression of the secondary flux by the time delay of the secondary signal introduced by the presence of an IGMF. From Eq. (\ref{eq:delay}) one finds that the field with the strength $B\sim 10^{-16}$~G suppresses the secondary flux by a factor $\kappa=(T_d/T)\gtrsim 10$ in the energy range below $E_\gamma=100$~GeV. 

To move beyond this order-of-magnitude estimate, we used the CRbeam code to model the secondary emission \citep{2016PhRvD..94b3007B,2023A&A...675A.132K}. We considered  an IGMF with the correlation length $\lambda_B=1$~Mpc $>l_e$ in the energy range of interest  \citep{2009PhRvD..80l3012N}.  The magnetic field was chosen to have a Kolmogorov power spectrum down to the wave number $k_{min}=2\pi/(5\lambda_B)$. We assumed that the source is switched on as a $\theta$ function at the start of the 16-year interval and calculated the average secondary signal spectrum over the 16-year time period from the start of the source activity. This secondary spectrum was added to the attenuated primary-source spectrum and the combined primary and secondary model spectrum was fit to the data. 

The time delay, $T_d$ is related to the angle of misalignment of the arrival direction of the secondary \gr\ $\theta$ as $T_d\sim \theta^2 D$, where $D$ is the distance to the \gr\ source \citep{2009PhRvD..80l3012N}. The time delay of the order of $20$~yr corresponds to the misalignment angle $\theta\sim 10''$, which is not resolvable with \gr\ telescopes. Given this fact, our model fitting is limited to the point-source spectrum. 

We compared the quality of fits of the data with the two alternative models (zero IGMF and infinite IGMF) using the $\chi^2$ minimization. Table \ref{tab:sources} provides the $\chi^2$ values of the fits of the data with the cut-off power-law model of the intrinsic spectrum (i.e., with a very large IGMF that completely suppresses the secondary flux), $\chi^2_\infty$ and with the zero-IGMF model, $\chi^2_0$. The $\chi^2$ values are quoted together with the number of degrees of freedom (the number of spectral data points minus 3, which is the number of parameters of the cut-off power-law model). We judged the source useful for the IGMF analysis if the $B=0$ model is found to be inconsistent with the data. Considering the magnetic-field strength as an additional parameter, we consider the model with zero magnetic field inconsistent with the data if $\Delta\chi^2=\chi^2_0-\chi^2_\infty>2.71,$ corresponding to the one-sided 95\% confidence level (CL). 

\begin{figure}
     \includegraphics[width=\columnwidth]{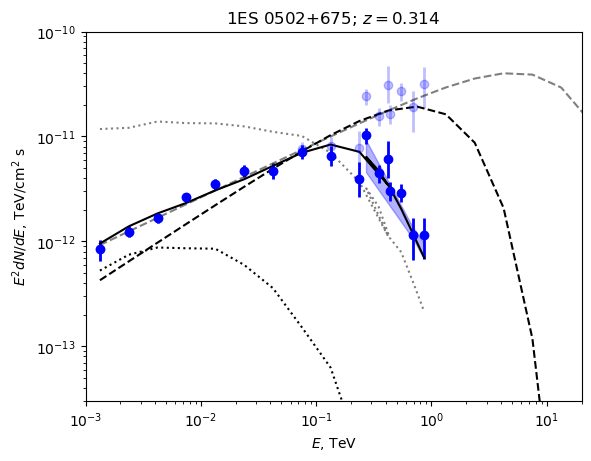}
   \caption{Spectral modeling for 1ES 0502+675. Blue data points show Fermi/LAT and VERITAS data. Lighter colored data points are corrected for the EBL absorption. The blue butterfly shows the measurement of the source spectrum from \cite{2019ApJ...885..150A}. The dashed gray line shows the best-fit intrinsic source cut-off power law expected in the case of a strong IGMF. The dotted gray line shows the secondary flux that is suppressed by the IGMF. The black curves show the model with minimal possible IGMF. The dashed line represents the intrinsic source spectrum, the dotted line the secondary flux, and the solid line the sum of the secondary and attenuated primary components.  }
    \label{fig:0502_nofield}
\end{figure}

\begin{figure*}
    \includegraphics[width=0.66\columnwidth]{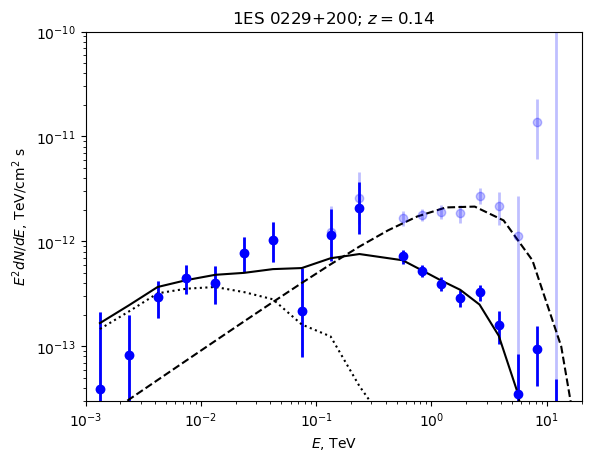}        
    \includegraphics[width=0.66\columnwidth]{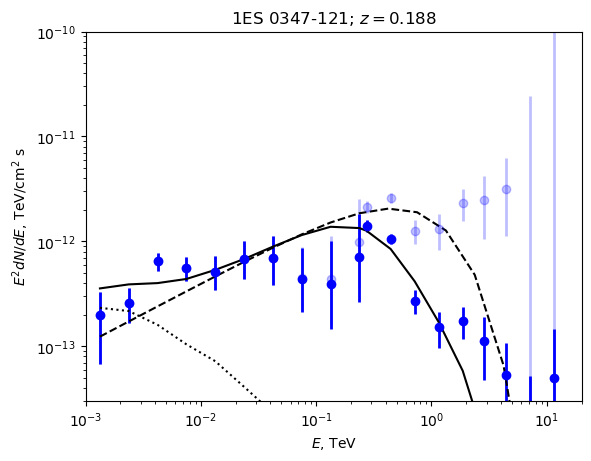}
    \includegraphics[width=0.66\columnwidth]{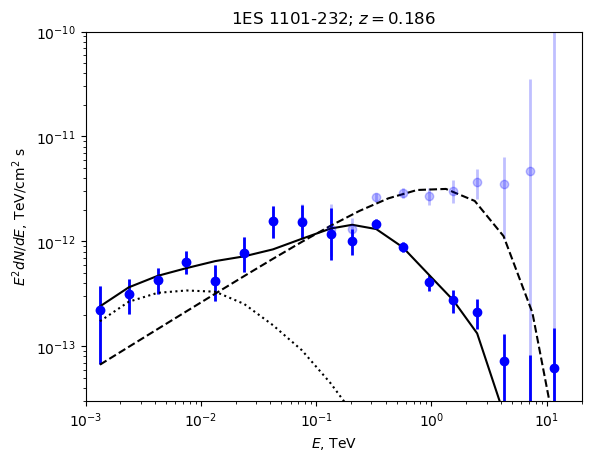}\\
    \includegraphics[width=0.66\columnwidth]{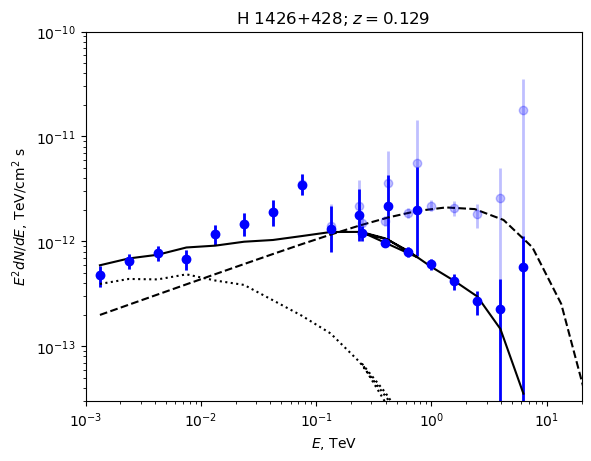}
    \includegraphics[width=0.66\columnwidth]{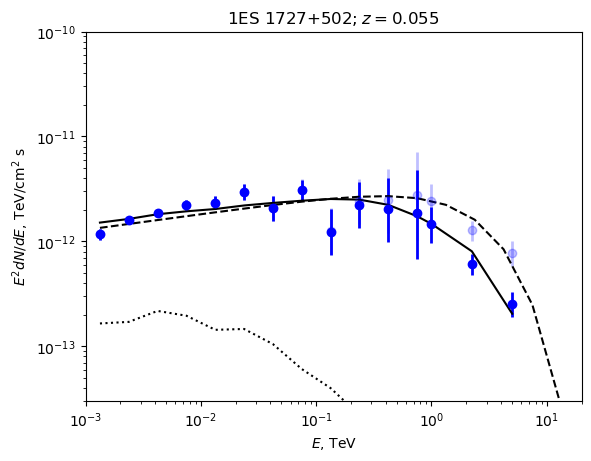}
    \includegraphics[width=0.66\columnwidth]{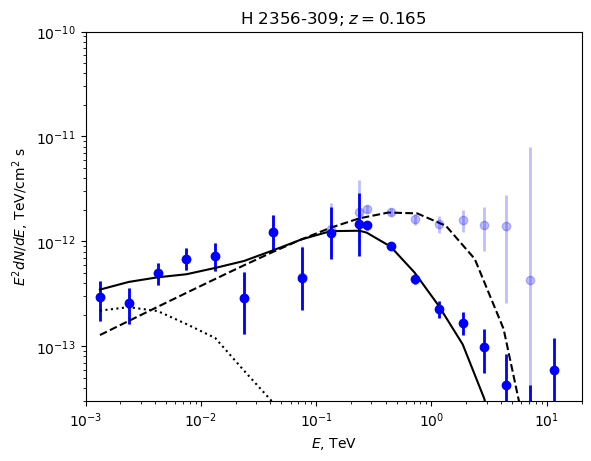}
   \caption{Spectra of sources that provide constraints on the IGMF. The data points are combinations of Fermi/LAT data with the IACT data for which references are listed in Table \ref{tab:sources}. The models correspond to the minimal possible field strength. The notations are the same as in Fig. \ref{fig:0502_nofield}.  }
    \label{fig:spectra}
\end{figure*}

\begin{figure}
     \includegraphics[width=\columnwidth]{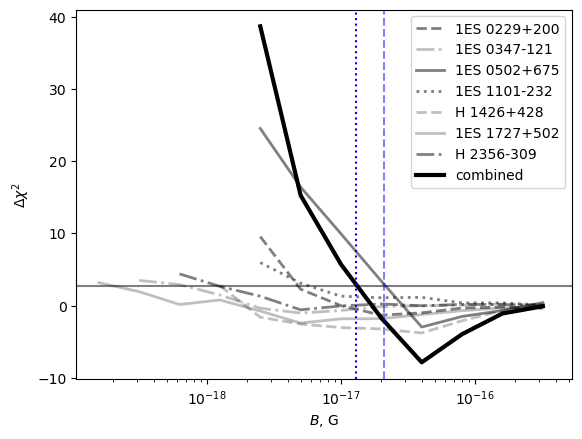}
   \caption{$\Delta \chi^2(B)$ profiles for sources that constrain the IGMF. The vertical dashed blue line shows the lower bound on $B$ from 1ES 0502+675, and the vertical dotted blue line shows the combined lower bound on the IGMF from all sources. Horizontal line shows $\Delta\chi^2=2.71$. }
    \label{fig:chi2}
\end{figure}

\subsection{Sources that do not yield constraints on the IGMF}

For most of the sources in Table \ref{tab:sources} we were able to find an acceptable fit to the Fermi/LAT $+$ IACT data using the $B=0$ model.  We show spectra of these sources in Fig. \ref{fig:candidates} in the appendix.  The spectrum of the primary source is shown by the dashed line and the secondary flux spectrum is shown by the dotted line. The sum of the attenuated primary source and the secondary source fluxes is shown by the solid line. As an example,  the spectrum of the first source, SHBL J001355.9-185406,  has a high-energy cut-off. This is the reason why the secondary flux expectation is low. The level of the secondary flux is a factor of two lower than the primary source flux at GeV energy, and a small adjustment of the slope of the primary source spectrum (compared to the model without the secondary flux component) is sufficient to make the model with the secondary flux consistent with the data.  In most of the other cases the cut-off in the spectrum is not constrained, but the quality of the VHE data is  insufficient for a good estimate of the secondary flux level.  It is clear that  an improvement in the quality of the VHE spectral measurements (e.g., via longer exposure with existing  IACTs or with the next-generation IACTs) could tighten the constraint on the intrinsic source-spectrum shape, in particular on the lower bound on $E_{cut}$. If the lower bound on $E_{cut}$ moves toward higher energy, the level of the secondary flux increases, and it can start to be too high to be compatible with Fermi/LAT measurements. This would make these sources useful for the IGMF analysis. If the improvement of the IACT data yielded a measurement of $E_{cut}$, precise predictions for the secondary flux level and spectra shape would become possible, further improving the precision of the IGMF analysis. 

One source that has previously been used to derive constraints on the IGMF by \cite{HESS:2023zwb}, namely PKS 0548-322, does not constrain the IGMF in our analysis, in spite of the fact that we used the same Fermi/LAT $+$ HESS data combination as \cite{HESS:2023zwb}. From Fig. \ref{fig:candidates} one can see that, contrary to what \cite{HESS:2023zwb} found, the spectrum of this source can be fit consistently with a model with zero IGMF. We explain the origin of this discrepancy in Appendix \ref{sec:comparison}.

\subsection{Sources that yield constraints on the IGMF}

There are seven sources for which the fit with zero IGMF is incompatible with the data: 1ES 0229+200, 1ES 0347-121, 1ES 0502+675, 1ES 1101-232, H 1426+428, 1ES 1727+502, and H 2356-309. Three of the sources---1ES 0229+200, 1ES 0347-121 and 1ES 1101-232---were used by \cite{2010Sci...328...73N} in the initial analysis that resulted in inference of existence of nonzero IGMF in the voids of the large-scale structure. H 2356-309 has been considered in the IGMF search by \citet{HESS:2023zwb}. Three other sources, 1ES 0502+675, 1ES 1727+502, and  H 1426+428 have not yet been considered in the IGMF analysis. 

\begin{table}
\caption{Limits on the IGMF from individual sources.}
\begin{tabular}{ll}
\hline
Name & $B_{min}/10^{-17}$~G \\
\hline
1ES 0229+200 & 0.5  \\
1ES 0347-121 & 0.07\\
1ES 0502+675 & 2.1\\
1ES 1101-232 & 0.6\\
H 1426+428 & 0.13 \\
1ES 1727+502 & 0.02\\
H 2356-309 & 0.12 \\
\hline
\end{tabular}
\label{tab:individual}
\end{table}

As an example, Fig. \ref{fig:0502_nofield} shows the spectrum of 1ES 0502+675, with two alternative models. The dashed gray line shows the model without the secondary flux (that with the secondary flux, represented by the dotted gray line, is suppressed by the IGMF). The black lines show the model with minimal possible magnetic field. In this case, the secondary flux (dotted black line) is only partially suppressed by the time delay due to the IGMF and is contributing to the point source flux, which is shown by the solid black line.  The minimal $B$ models for the spectra of the other six sources are shown in Fig. \ref{fig:spectra}.

Figure \ref{fig:chi2} shows the $\Delta\chi^2=\chi^2(B|E_{cut},\Gamma,N)-\chi^2_\infty$ of the model fits as a function of $B$, with the intrinsic spectrum parameters, $E_{cut}, \Gamma, N,$ fixed to their best-fit values for each $B$.  Following \cite{1976ApJ...210..642A}, we assumed that $\chi^2(B)=\chi^2(B|E_{cut},\Gamma,N)$ is distributed as $\chi^2_1$ with one degree of freedom and estimated a one-sided 95\% lower bound on $B$ by requiring $\Delta\chi^2<2.71$ for the models with acceptable  $B$. In this sense, our approach is equivalent to that of \cite{HESS:2023zwb}, where the same principle was applied to the profile likelihood as a function of $B$.

The strongest constraint on the IGMF, 
\begin{equation}
\label{eq:bound}
    B_{min,0502}>2.1\times 10^{-17}\mbox{ G}
,\end{equation}
stems from the spectral fit to the data of 1ES 0502+675 (Fig. \ref{fig:chi2}). \citet{2025arXiv250608497N} noticed that this is the second-brightest VHE-detected extreme blazar after Mrk 501. It has a hard \gr\ spectrum extending up to 1 TeV and as such it is a better candidate for constraining the IGMF. Our analysis confirms this expectation.  The
IACT data considered in our analysis are from  \citet{Orr:2013yma}. From Fig. \ref{fig:0502_nofield} one can see that the combined Fermi/LAT and VERITAS source spectrum does not show a signature of high-energy cut-off, once corrected for the EBL. The spectrum is hard, with the slope $\Gamma$ consistent with the hardest possible slope of inverse Compton emission (1.5, see Table \ref{tab:sources}). The fit with the smallest possible IGMF features an even harder intrinsic power-law slope: $\Gamma=1.3$. 
Table \ref{tab:individual} lists the constraints on $B$ imposed by the data on the other six sources. Fig. \ref{fig:chi2} shows the $\Delta\chi^2$ profiles for all seven sources.

Remarkably, the source  1ES 0229+200, which was found to yield strongest constraints in the HESS and Fermi/LAT analysis by \citet{HESS:2023zwb}, is found to yield an order-of-magnitude weaker bound in spite of the fact that the same HESS data were used in our analysis.  We explain in Appendix \ref{sec:comparison} that this difference stems from the mismatches in the modeling of the secondary flux in the two analyses. Our secondary flux model matches that of \citet{2023A&A...670A.145A}. The spectrum of the model with the smallest possible IGMF in this case is shown in the left column of the top panel of Fig. \ref{fig:spectra}.  

 Our bound on the IGMF from 1ES 0229+200 data  is broadly consistent with that derived by \citet{2023A&A...670A.145A} based on a combination of Fermi/LAT, MAGIC, VERITAS, and HESS data.  
 \cite{2023A&A...670A.145A} noticed that 1ES 0229+200 is variable on the timescale of 20~years of historical observations with IACTs. This variability can be taken into account self-consistently by modeling the light curve of the delayed secondary flux using the knowledge of the primary source light curve. In the simplified version of our analysis, we did not consider the details of the primary source variability and applied  a different assumed activity period of 16 years. Moreover, the best-fit energy cut-off for the IGMF bound inferred from 1ES 0229+200 is $E_{cut}\sim 2.5$ TeV (see Table \ref{tab:best_fit_params_B_field}), which is somewhat lower than the one used in \cite{2023A&A...670A.145A} where $E_{cut}=7$ TeV; the latter can lead to a slightly larger suppression of the secondary cascade. This explains the fact that our bound from 1ES 0229+200 data is somewhat weaker than that of  \citet{2023A&A...670A.145A}.
 
 A bound similar to that  from the 1ES 0229+200 data is provided by the 1ES 1101-232 data (see Fig. \ref{fig:chi2} and Table \ref{tab:individual}).  Also, this bound is significantly lower than the bound derived by \cite{HESS:2023zwb}, with the discrepancy stemming from the modeling of the secondary signal. Four more sources---1ES 0347-121, H 1426-428, 1ES 1727+502, and H 2356-309---provide lower bounds on the IGMF that are  much weaker than that imposed by the 1ES 0502+675 modeling. The weakness of the lower bound from 1ES 0347-121 can be partially understood as stemming from the poor quality of the spectral fits. Table \ref{tab:sources} indicates that the $\chi^2_\infty$ of the fit of the intrinsic source spectrum with a cut-off power-law model is 29.1 for 16 degrees of freedom (number of spectral measurement points minus the number of the model parameters). This means that the cut-off power-law model is actually inconsistent with the data at a $>97\%$ confidence level. The spectrum of the source is shown in the top right panel of Fig. \ref{fig:spectra}. One can see that the spectral data show large deviations from the power-law behavior, and HESS data do not match the highest energy data points of Fermi/LAT well. This can be because of variability of the source: Fermi/LAT measures the time-averaged source flux, while HESS performs snapshot observations on an irregular basis, which can catch some activity episodes and miss quiescent periods. In fact, the fit favors the model with a magnetic field somewhat below $10^{-17}$~G, as can be seen from the $\Delta \chi^2$ plot for this source in Fig. \ref{fig:chi2}. In this case the "hump" in the spectrum provided by the secondary flux component fits one of the deviations from the power law in the data. The same is true for H 1426+428. The source spectrum is shown in the right column in the second row of Fig. \ref{fig:spectra}. The fit favors certain values of $B$  for which the secondary flux component fits a hump in the Fermi/LAT spectrum. 

Following \cite{HESS:2023zwb}, we also considered the combination of IGMF lower bounds by adding the $\Delta\chi^2$ for all the seven sources.  Similar to the single source analysis, for each value of $B$, we find the best-fit intrinsic spectral parameters  $E_{cut,i},\Gamma_i, N_i$ for each $i$th source separately, but the value of $B$ is common for all sources. Adopting the approach of \cite{1976ApJ...210..642A} for this case, we assumed that the cumulative $\chi^2(B|E_{cut,1},\Gamma_{cut,1},N_1,...E_{cut,7},\Gamma_7,N_7)$ also follows the $\chi^2$ statistics with one degree of freedom. The combination of data on all sources slightly relaxes the lower bound on the IGMF, down to  
\begin{equation}
    B_{min,all}>1.3\times 10^{-17}\mbox{ G.}
\end{equation}
This is explained by the fact that the bound from 1ES 0502+675 data is significantly stronger than the bounds from all other sources. 

\begin{figure}
     \includegraphics[width=\columnwidth]{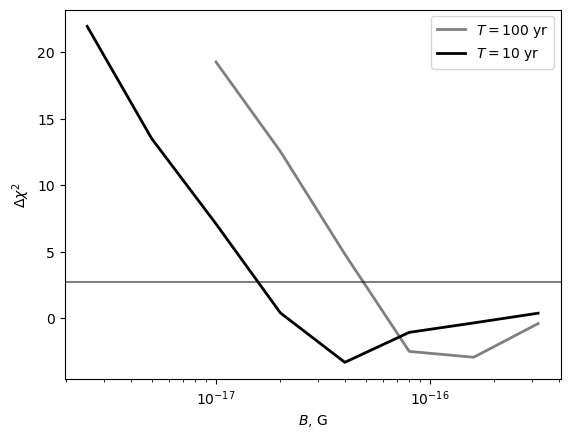}
   \caption{Comparison of $\Delta \chi^2(B)$ profiles for 1ES 0502+675 for two different assumptions about the source-activity period. }
    \label{fig:chi2_comparison}
\end{figure}

One can also notice in Fig. \ref{fig:chi2} that the cumulative $\chi^2$ profile has a minimum at approximately $B\simeq 4\times 10^{-17}$~G, which is driven by the minimum in the $\chi^2$ profile of 1ES 0502+576 and is further strengthened by the mild minima in the $\chi^2$ profiles of other sources. We did not consider this minimum of the $\chi^2$ profile as a measurement of IGMF, but rather as an artifact of our modeling approach. It is related to the (unrealistic) assumption that all seven sources considered in Fig. \ref{fig:chi2} had just been switched on approximately ten years ago at the start of observations in the VHE \gr\ band. The IGMF with the strength $B\simeq 4\times 10^{-17}$~G induces the time delay that is on the order of 16~yr in the $E\sim 10$~GeV energy
range (see Eq. (\ref{eq:delay})). The delay timescale much longer than 10~yr in the GeV range produces a significant suppression of the secondary flux at GeV, and a much more modest suppression in the 10~GeV range. This induces a curvature in  the cascade spectrum at around a 10~GeV energy level, as can be seen from Fig. \ref{fig:0502_nofield}. The statistics of the Fermi/LAT signal in the 10~GeV range is sufficiently high that small deviations from the power-law behavior can be observed. The curved spectrum of the secondary emission component helps in the modeling of the deviation of the spectrum from the power law in this energy range. To demonstrate this effect explicitly, in Fig. \ref{fig:chi2_comparison} we show a comparison of the $\Delta\chi^2$ profiles for 1ES 0502+675 for two different assumptions about the duration of the source activity cycle: $T=10$~yr and $T=100$~yr. One can see that the minimum of the  $\chi^2$ shifts toward higher magnetic-field strength, because a somewhat stronger field is required to achieve significant suppression of the secondary flux in the GeV range while keeping only a moderate suppression at 10~GeV. 

\section{Discussion and conclusions}

We used the set of extreme blazars detected in the VHE band by both Fermi/LAT and IACTs \citep{2025arXiv250608497N} to revise the lower bound on magnetic fields in the voids of the large-scale structure. This bound  stems from the observation of the effect of suppression of the secondary \gr\ flux from electron-positron pairs deposited in the voids as a result of interactions of VHE \gr s with the optical and infrared photons of the EBL. Similarly to  \cite{2023A&A...670A.145A} and \cite{HESS:2023zwb}, we considered the assumption that sources were only "switched on" at the start of observations in the VHE band as "conservative" (perhaps "over-conservative"), and the secondary emission was only suppressed because of the time delay of secondary flux caused by deflections of electrons and positrons in the IGMF. 

In the specific case of the blazar 1ES 0229+200, considered by \cite{2023A&A...670A.145A}, our bound is somewhat weaker than that derived by \cite{2023A&A...670A.145A}, because we did not follow the details of source variability over the 20 years of source observations in the VHE band, and we assumed that the source luminosity is represented by a $\theta$-function-like switch-on followed by the constant luminosity on the 16-year timescale (representative of the time-averaged source luminosity). Comparing our IGMF limit with that derived by \cite{HESS:2023zwb}, we find a weaker lower bound, even though our modeling of source "switch-on" is the same as that of \citet{HESS:2023zwb}. We argue that the mismatch between our results and those of \cite{2023A&A...670A.145A} and the results of  \cite{HESS:2023zwb} stems from a discrepancy in the modeling of the secondary flux (see Appendix \ref{sec:comparison} for details). 

Even though our IGMF limit from the data on 1ES 0229+200 is weaker than that of \cite{2023A&A...670A.145A}, we identified another source, 1ES 0502+675, that provides a tighter lower bound on IGMF at the$B>2.1\times 10^{-17}$~G level. We checked that combination of the data on all sources that yield constraints on IGMF (seven in total) slightly worsens the lower bound on IGMF down to  $B>1.3\times 10^{-17}$~G. It is surprising that  1ES 0502+675 has not been considered before in the IGMF search, and no refereed journal publication\footnote{A power-law spectral fit for this source is given in \cite{2019ApJ...885..150A}. We show this fit in Fig. \ref{fig:0502_nofield}.} about the source spectrum has appeared since the detection of the source back in 2009 by VERITAS \citep{2009ATel.2301....1O}. We were only able to find a conference-proceedings report on the source spectrum. Given that the source provides the tightest constraints on IGMF in our analysis, we believe that a better characterization of its TeV band spectrum with longer IACT exposures would be useful. The source is very bright (the second-brightest VHE extreme blazar after Mrk 501, as noted in \citet{2025arXiv250608497N}), so detailed characterization of its spectrum in the TeV band should be possible with current-generation IACTs. Such measurements could better constrain the energy cut-off and possibly extend the spectrum in the energy range above 1~TeV, which would allow for a further tightening of the conservative lower bound on the IGMF.

A potential caveat to these constraints is the relaxation of the electron-positron pair beam via collective plasma instabilities, which could dissipate the beam energy into the intergalactic medium rather than producing secondary $\gamma$-rays. Several studies have argued that the cooling time due to interactions between the relativistic pair beams and the intergalactic medium can be shorter than the inverse Compton cooling time translating to plasma instabilities that can reduce the secondary cascade flux \citep{Broderick2012,Broderick2018,Vafin:2019} and lead to a smaller inferred lower bound on the IGMFs of blazars (see, e.g., \citet{DeySigl2025} for the case of 1ES 0229+200). Some models suggest that large-scale magnetic fields might suppress the growth of these instabilities \citep{Alawashra:2022all,Alawashra:2024ndp}. The results of the laboratory experiment in \cite{Arrowsmith:2025apl} demonstrated that the instability growth rates are heavily suppressed unless the particle beam is perfectly collimated or monochromatic. These ideal conditions are usually not met by blazar-induced pairs. Hence, inferring lower limits on the IGMF strength from the non-observation of the GeV cascade of blazars remains a robust method.

In our analysis we focused on an IGMF with a long correlation length of $\lambda_B>l_e$. Nevertheless, we can extrapolate the results to the case of short correlation-length fields using prescriptions of \citet{2009PhRvD..80l3012N}. The lower bound is independent of $\lambda_B$ as long as $\lambda_B>l_e$ and it scales as $\lambda_B^{-1/2}$ in the $\lambda_B<l_e$ range. The estimate of $l_e$ in our analysis is identical to that of \cite{2023A&A...670A.145A}, $l_e\sim 0.1$~Mpc, so our bound for the short correlation length field is 
\begin{equation}
    B\gtrsim 2.1\times 10^{-17}\mbox{ G}\left\{ 
    \begin{array}{ll}
    \left[\frac{\displaystyle \lambda_B}{\displaystyle 0.1\mbox{ Mpc}}\right]^{-1/2}, & \lambda_B<0.1\mbox{ Mpc}\\
    1, & \lambda_B\ge 0.1\mbox{ Mpc}
    \end{array}
    \right.
.\end{equation}
The short-correlation-length field bound is particularly interesting for the cosmological IGMF that may originate from cosmological phase transitions (electroweak, quark confinement) \citep{2013A&ARv..21...62D}. For such fields the correlation length and strength are related by the "largest processed eddy" condition. There exist alternative estimates of the largest processed eddy scale, with the estimate of \citet{2004PhRvD..70l3003B} being $\lambda\sim 1~[B/10^{-8.5}\mbox{ G}]$~Mpc. \citet{2023NatCo..14.7523H} and \citet{2024A&A...687A.186B} estimated it to be two to four orders of magnitude shorter. This gives a lower bound of $\sim 10^{-14}$~G on the cosmological magnetic field under the assumption of  \citet{2004PhRvD..70l3003B} largest processed eddy scaling and up to two orders of magnitude stronger assuming the  \citet{2023NatCo..14.7523H} largest processed eddy scaling. 
The cosmological relic magnetic field with the strength close to this lower bound can, for example, be a helical magnetic field that was responsible for the generation of baryon asymmetry of the Universe during the electroweak cross-over \citep{2016PhRvD..94f3501K,2025arXiv250407937B}, or it can be a non-helical field generated at quark confinement first-order phase transition \citep{1997PhRvD..55.4582S}.

The field at the level  close to the lower bound derived above should be detectable using multiple techniques: through detection of delayed "pair-echo" emission in the signals of AGNs \citep{2023A&A...670A.145A} and \gr~bursts \citep{2024A&A...688A..57M,2011MNRAS.410.2741T}, or through the detection of extended emission around AGNs \citep{2009PhRvD..80l3012N,2018ApJS..237...32A,HESS:2023zwb}. To identify the best targets for such a search, it is necessary to improve our knowledge of the intrinsic spectra of sources from Table \ref{tab:sources}. For most of them, no certain predictions for the secondary signal flux can be made because of uncertainty of the high-energy cut-offs in the spectra. This can be improved using the new generation of IACTs; that is, the Cherenkov Telescope Array Observatory (CTAO) \citep{2019scta.book.....C} and the Large Array of imaging atmospheric Cherenkov Telescope (LACT) \citep{2025ChPhC..49c5001Z}, which have better sensitivity in the multi-TeV energy range. The information on the time-average intrinsic fluxes of the sources and on the presence of high-energy cut-offs in their spectra can also be provided by air-shower arrays, such as the already-operating Large High Altitude Air Shower Observatory (LHAASO) \citep{2024ApJS..271...25C} and the planned Southern Wide-field Gamma-ray Observatory (SWGO) \citep{2019BAAS...51g.109H}. 

\begin{acknowledgements}
We would like to thank J. Biteau for clarifying discussions as well as the anonymous referee for their useful suggestions to improve the manuscript. This work has been supported in part by the French National Research Agency (ANR) grant ANR-24-CE31-4686. 
\end{acknowledgements}

\bibliographystyle{aa}
\bibliography{refs_aa}

\begin{appendix}

\section{Modeling of the secondary flux and comparison with \citet{HESS:2023zwb}}
\label{sec:comparison}

\begin{figure}
    \includegraphics[width=\columnwidth]{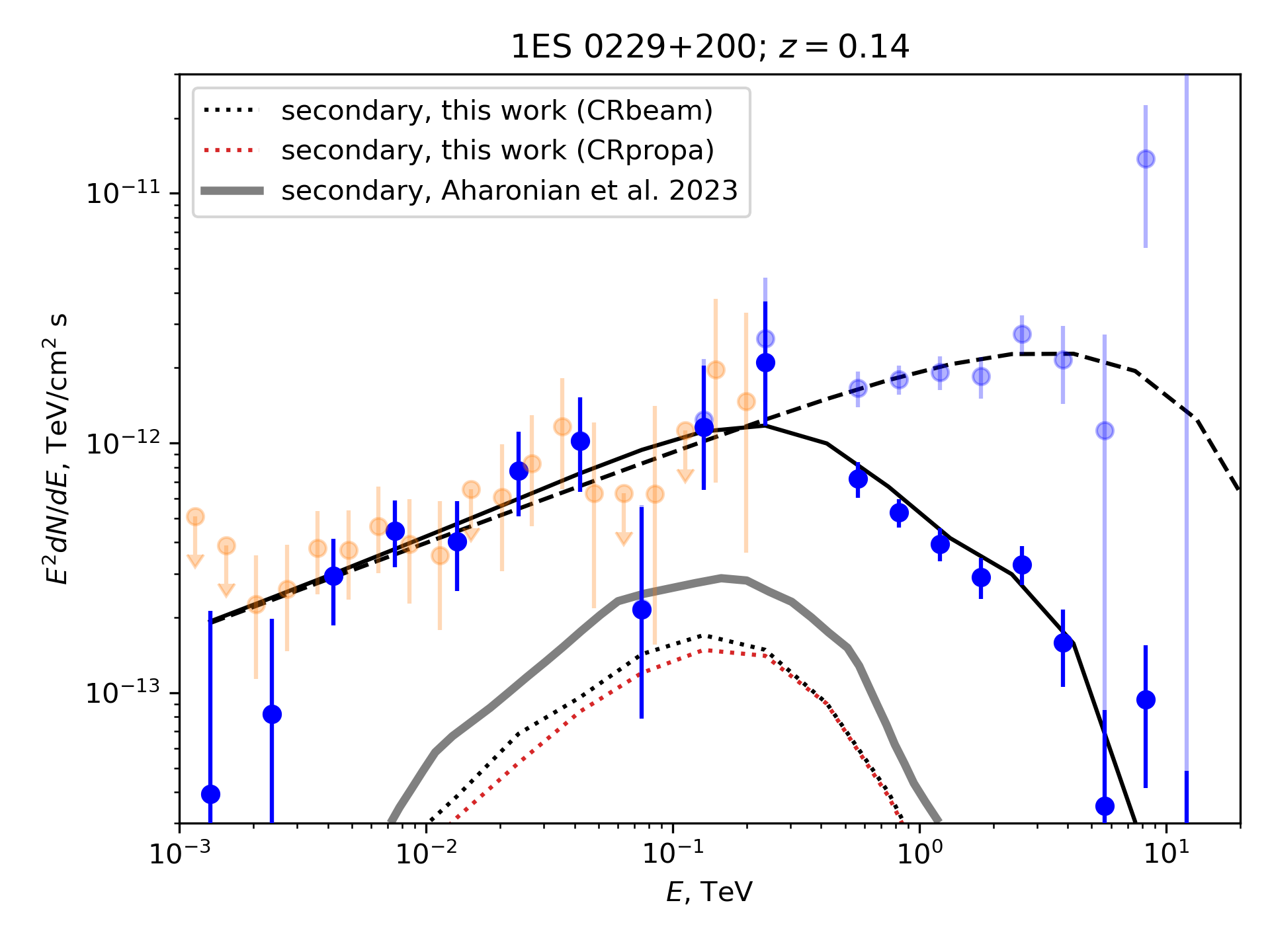}
    \caption{Comparison  of the secondary flux from \citet{HESS:2023zwb} with our secondary flux model for the same parameters (magnetic field $B=3.2\times 10^{-16}$~G, activity time $T=10$~yr, source 1ES 0229+200). Results from CRbeam (black dotted line) and CRpropa (red dotted line) are both present. Orange data points show the result of Fermi/LAT likelihood analysis by \cite{HESS:2023zwb}.}
    \label{fig:comparison}
\end{figure}

We compute the secondary flux using the CRbeam code \citep{2016PhRvD..94b3007B,2023A&A...675A.132K}. We simulate a beam of photons in the energy range between 10~GeV and 30~TeV, with $dN/dE\propto E^{-1}$ spectrum. For each secondary photon produced following the $e^+e^-$ pair production followed by the inverse Compton scattering, we keep a memory of the primary photon. This allows us to calculate the secondary flux for any primary photon spectrum using single simulation of the $E^{-1}$ spectrum, by simply weighting the secondary photons according to the energy of the primary photon. We propagate primary and secondary particles until they cross a sphere of the radius $D$ equal to the distance to the source of interest. For each photon crossing the sphere, we calculate the angle between the direction of the photon and normal to the sphere, which corresponds to the angle of misalignment of the arrival direction of the secondary photons. We also notice the total path from the source for each photon crossing the sphere. This path, divided by the speed of light, gives the arrival time of all photons. For completeness, we run the same configuration considered in CRbeam with CRpropa \citep{2022JCAP...09..035A} and calculate the corresponding secondary flux following the procedure described above.

Fig. \ref{fig:comparison} shows a comparison of our calculation of the secondary flux for $T=10$~yr activity time of 1ES 0229+200, for the magnetic field $B=3.2\times 10^{-16}$~G with that of \cite{HESS:2023zwb}\footnote{The original secondary flux curve has been extracted from slide 28 of the following presentation: https://indico.ict.inaf.it/event/3156/contributions/21372/attachments/9760/20166/MMeyer\_IGMF\_Trieste\_2025.pdf}. Our estimate of the secondary flux is by up to a factor of two lower than that of \cite{HESS:2023zwb}. This difference does not stem from the discrepancies between CRbeam code and CRpropa code used by \cite{HESS:2023zwb} as shown by Fig. A.1, which is consistent with \cite{2023A&A...675A.132K} where  the two codes have been found to agree within the redshift range of interest. CRpropa code was also used in the modeling of the secondary flux by \cite{2023A&A...670A.145A}, with which our calculations agree.

Our analysis is based on Fermi/LAT spectra extracted using the aperture photometry technique, as compared to the likelihood analysis used by \cite{HESS:2023zwb}. To estimate the source flux in each energy bin, we use the knowledge of the 68\% and 95\% containment radii of the Fermi/LAT point spread function. If the total source signal in the energy bin is $S$ and the background per square degree is $b$, the numbers of photon counts in the 68\% and 95\% containment circles are
\begin{eqnarray}
    &&S_{68}=0.68 S+\pi (0.68^\circ)^2 b\nonumber\\
    &&S_{95}=0.95 S+\pi (0.95^\circ)^2 b
\end{eqnarray}
so that the signal can be found from the measurements of $S_{68}$ and $S_{95}$:
\begin{equation}
    S=\frac{0.95^2S_{68}-0.68^2S_{95}}{0.95^2-0.68^2}
\end{equation}
We use this estimate in the energy range below 50~GeV, where the background level is high. In the energy range above 50~GeV, the sources are mostly detected in the background-free regime and we assume that $S=S_{95}/0.95$. The aperture photometry approach has an advantage compared to the likelihood analysis approach in the "uncrowded" sky regions where there are no sources with overlapping signals. Contrary to the likelihood estimate,  it does not depend on the sky model that has many parameters (of the Galactic diffuse emission, isotropic diffuse emission, distribution of point sources in the region of interest, choice of the spectral models and spectral parameters of those sources). We have checked that our aperture photometry spectra are consistent with those of the likelihood analysis by \citet{HESS:2023zwb}. Fig. \ref{fig:comparison} shows a comparison of the aperture photometry and likelihood analysis for 1ES 0229+200.

\onecolumn
\renewcommand{\arraystretch}{1.25}
\begin{longtable}{cccccc}
\caption{Best-fit spectral parameters of the sources listed in Table \ref{tab:individual} for different IGMF strengths.}\\
\hline
\hline
& & & & & \\
& $B$ [G] & $N_0$ [$10^{-12}$ TeV$^{-1}$cm$^{-2}$s$^{-1}$] & $\Gamma$  & $E_{cut}$ [TeV] &\\
& & & & & \\
\hline
\hline
\multicolumn{6}{c}{1ES 0229+200} \\
\hline
& $6.00\times10^{-19}$ & $3.63^{+0.74}_{-0.61}$ & $1.17^{+0.12}_{-0.16}$ & $1.78^{+0.41}_{-0.30}$ &\\
& $1.20\times10^{-18}$ & $3.55^{+0.72}_{-0.53}$ & $1.17^{+0.12}_{-0.17}$ & $1.86^{+0.43}_{-0.35}$ &\\
& $2.39\times10^{-18}$ & $3.47^{+0.60}_{-0.52}$ & $1.16^{+0.13}_{-0.16}$ & $2.09^{+0.48}_{-0.39}$ &\\
& $4.76\times10^{-18}$ & $3.24^{+0.56}_{-0.49}$ & $1.18^{+0.13}_{-0.17}$ & $2.45^{+0.64}_{-0.45}$ &\\
& $9.50\times10^{-18}$ & $2.95^{+0.52}_{-0.44}$ & $1.25^{+0.12}_{-0.15}$ & $3.16^{+0.91}_{-0.65}$ &\\
& $1.89\times10^{-17}$ & $2.69^{+0.47}_{-0.40}$ & $1.35^{+0.10}_{-0.13}$ & $4.27^{+1.62}_{-1.03}$ &\\
& $3.78\times10^{-17}$ & $2.51^{+0.44}_{-0.42}$ & $1.44^{+0.08}_{-0.11}$ & $5.62^{+3.29}_{-1.73}$ &\\
& $7.54\times10^{-17}$ & $2.34^{+0.41}_{-0.43}$ & $1.52^{+0.06}_{-0.08}$ & $7.41^{+6.08}_{-2.73}$ &\\
& $1.50\times10^{-16}$ & $2.45^{+0.50}_{-0.54}$ & $1.56^{+0.06}_{-0.06}$ & $7.24^{+8.98}_{-2.77}$ &\\
& $3.00\times10^{-16}$ & $2.19^{+0.50}_{-0.64}$ & $1.60^{+0.07}_{-0.06}$ & $10.72^{+20.90}_{-5.35}$ &\\
\hline
\multicolumn{6}{c}{1ES 0347-121} \\
\hline
& $6.00\times10^{-19}$ & $7.08^{+1.63}_{-1.33}$ & $1.36^{+0.07}_{-0.08}$ & $0.83^{+0.22}_{-0.17}$ &\\
& $1.20\times10^{-18}$ & $6.46^{+1.67}_{-1.21}$ & $1.39^{+0.07}_{-0.08}$ & $0.91^{+0.26}_{-0.20}$ &\\
& $2.39\times10^{-18}$ & $5.75^{+1.33}_{-1.08}$ & $1.40^{+0.07}_{-0.08}$ & $1.12^{+0.36}_{-0.27}$ &\\
& $4.76\times10^{-18}$ & $5.13^{+1.33}_{-0.96}$ & $1.45^{+0.06}_{-0.07}$ & $1.29^{+0.53}_{-0.38}$ &\\
& $9.50\times10^{-18}$ & $4.68^{+1.21}_{-0.88}$ & $1.50^{+0.06}_{-0.06}$ & $1.48^{+0.76}_{-0.46}$ &\\
& $1.89\times10^{-17}$ & $4.47^{+1.29}_{-0.92}$ & $1.54^{+0.05}_{-0.06}$ & $1.58^{+0.99}_{-0.54}$ &\\
& $3.78\times10^{-17}$ & $4.37^{+1.26}_{-0.98}$ & $1.57^{+0.05}_{-0.06}$ & $1.55^{+1.21}_{-0.55}$ &\\
& $7.54\times10^{-17}$ & $3.98^{+1.15}_{-0.89}$ & $1.59^{+0.06}_{-0.06}$ & $1.95^{+2.12}_{-0.77}$ &\\
& $1.50\times10^{-16}$ & $3.89^{+1.24}_{-0.94}$ & $1.60^{+0.06}_{-0.07}$ & $2.04^{+2.13}_{-0.92}$ &\\
& $3.00\times10^{-16}$ & $3.80^{+1.21}_{-0.92}$ & $1.61^{+0.07}_{-0.07}$ & $2.24^{+2.55}_{-1.01}$ &\\
\hline
\multicolumn{6}{c}{1ES 0502+675} \\
\hline
& $6.00\times10^{-19}$ & $93.33^{+16.32}_{-13.90}$ & $1.23^{+0.04}_{-0.04}$ & $0.35^{+0.03}_{-0.03}$ &\\
& $1.20\times10^{-18}$ & $91.20^{+15.95}_{-13.58}$ & $1.23^{+0.04}_{-0.04}$ & $0.38^{+0.04}_{-0.03}$ &\\
& $2.39\times10^{-18}$ & $89.13^{+15.58}_{-13.27}$ & $1.23^{+0.04}_{-0.04}$ & $0.43^{+0.04}_{-0.04}$ &\\
& $4.76\times10^{-18}$ & $83.18^{+14.54}_{-12.39}$ & $1.23^{+0.04}_{-0.05}$ & $0.52^{+0.06}_{-0.06}$ &\\
& $9.50\times10^{-18}$ & $75.86^{+13.27}_{-11.30}$ & $1.22^{+0.05}_{-0.05}$ & $0.68^{+0.08}_{-0.07}$ &\\
& $1.89\times10^{-17}$ & $63.10^{+9.34}_{-9.40}$ & $1.23^{+0.06}_{-0.06}$ & $0.98^{+0.14}_{-0.15}$ &\\
& $3.78\times10^{-17}$ & $56.23^{+9.83}_{-8.37}$ & $1.19^{+0.09}_{-0.11}$ & $1.55^{+0.23}_{-0.23}$ &\\
& $7.54\times10^{-17}$ & $44.67^{+6.62}_{-6.65}$ & $1.32^{+0.05}_{-0.09}$ & $2.09^{+0.54}_{-0.50}$ &\\
& $1.50\times10^{-16}$ & $39.81^{+6.96}_{-5.93}$ & $1.40^{+0.04}_{-0.04}$ & $2.29^{+0.80}_{-0.67}$ &\\
& $3.00\times10^{-16}$ & $37.15^{+6.50}_{-6.25}$ & $1.44^{+0.03}_{-0.03}$ & $2.63^{+1.35}_{-0.93}$ &\\
\hline 
\multicolumn{6}{c}{1ES 1101-232} \\
\hline
& $6.00\times10^{-19}$ & $10.00^{+2.02}_{-1.49}$ & $1.20^{+0.09}_{-0.11}$ & $0.87^{+0.13}_{-0.11}$ &\\
& $1.20\times10^{-18}$ & $9.12^{+1.84}_{-1.36}$ & $1.22^{+0.08}_{-0.10}$ & $0.98^{+0.17}_{-0.15}$ &\\
& $2.39\times10^{-18}$ & $8.71^{+1.52}_{-1.30}$ & $1.22^{+0.09}_{-0.10}$ & $1.12^{+0.20}_{-0.17}$ &\\
& $4.76\times10^{-18}$ & $7.94^{+1.39}_{-1.18}$ & $1.23^{+0.09}_{-0.11}$ & $1.35^{+0.24}_{-0.20}$ &\\
& $9.50\times10^{-18}$ & $6.92^{+1.02}_{-0.89}$ & $1.28^{+0.09}_{-0.11}$ & $1.74^{+0.35}_{-0.29}$ &\\
& $1.89\times10^{-17}$ & $6.03^{+0.89}_{-0.78}$ & $1.35^{+0.08}_{-0.10}$ & $2.29^{+0.59}_{-0.47}$ &\\
& $3.78\times10^{-17}$ & $5.37^{+0.80}_{-0.80}$ & $1.45^{+0.06}_{-0.07}$ & $2.82^{+1.26}_{-0.78}$ &\\
& $7.54\times10^{-17}$ & $4.79^{+0.84}_{-0.81}$ & $1.50^{+0.05}_{-0.06}$ & $3.72^{+2.59}_{-1.26}$ &\\
& $1.50\times10^{-16}$ & $4.47^{+0.78}_{-1.00}$ & $1.54^{+0.05}_{-0.05}$ & $4.47^{+5.08}_{-1.78}$ &\\
& $3.00\times10^{-16}$ & $4.27^{+0.86}_{-1.51}$ & $1.57^{+0.08}_{-0.05}$ & $5.37^{+16.01}_{-2.42}$ &\\
\hline
\multicolumn{6}{c}{H 1426+428} \\
\hline
& $6.00\times10^{-19}$ & $3.09^{+0.38}_{-0.34}$ & $1.58^{+0.05}_{-0.06}$ & $3.16^{+1.10}_{-0.71}$ &\\
& $1.20\times10^{-18}$ & $2.88^{+0.35}_{-0.37}$ & $1.58^{+0.06}_{-0.07}$ & $3.98^{+1.78}_{-1.03}$ &\\
& $2.39\times10^{-18}$ & $2.69^{+0.40}_{-0.40}$ & $1.60^{+0.06}_{-0.07}$ & $5.62^{+3.92}_{-1.99}$ &\\
& $4.76\times10^{-18}$ & $2.45^{+0.43}_{-0.79}$ & $1.66^{+0.04}_{-0.07}$ & $7.94^{+20.87}_{-3.38}$ &\\
& $9.50\times10^{-18}$ & $2.24^{+0.45}_{-0.62}$ & $1.73^{+0.05}_{-0.04}$ & $10.23^{+21.36}_{-5.10}$ &\\
& $1.89\times10^{-17}$ & $2.19^{+0.44}_{-0.49}$ & $1.77^{+0.05}_{-0.03}$ & $12.30^{+19.34}_{-6.27}$ &\\
& $3.78\times10^{-17}$ & $2.19^{+0.38}_{-0.49}$ & $1.79^{+0.05}_{-0.03}$ & $12.59^{+19.03}_{-6.42}$ &\\
& $7.54\times10^{-17}$ & $2.29^{+0.40}_{-0.51}$ & $1.79^{+0.05}_{-0.03}$ & $10.72^{+20.90}_{-5.35}$ &\\
& $1.50\times10^{-16}$ & $2.40^{+0.36}_{-0.45}$ & $1.79^{+0.04}_{-0.03}$ & $8.71^{+13.20}_{-4.03}$ &\\
& $3.00\times10^{-16}$ & $2.45^{+0.36}_{-0.41}$ & $1.79^{+0.04}_{-0.03}$ & $7.94^{+11.12}_{-3.48}$ &\\
\hline
\multicolumn{5}{c}{1ES 1727+502} \\
\hline
& $6.00\times10^{-19}$ & $4.37^{+1.01}_{-0.82}$ & $1.81^{+0.04}_{-0.04}$ & $2.19^{+0.63}_{-0.45}$ &\\
& $1.20\times10^{-18}$ & $4.27^{+0.98}_{-0.80}$ & $1.82^{+0.04}_{-0.04}$ & $2.24^{+0.71}_{-0.46}$ &\\
& $2.39\times10^{-18}$ & $4.17^{+0.96}_{-0.78}$ & $1.83^{+0.04}_{-0.04}$ & $2.29^{+0.66}_{-0.47}$ &\\
& $4.76\times10^{-18}$ & $4.27^{+0.86}_{-0.80}$ & $1.83^{+0.04}_{-0.04}$ & $2.29^{+0.66}_{-0.47}$ &\\
& $9.50\times10^{-18}$ & $4.07^{+0.82}_{-0.76}$ & $1.84^{+0.04}_{-0.04}$ & $2.34^{+0.68}_{-0.48}$ &\\
& $1.89\times10^{-17}$ & $4.17^{+0.84}_{-0.70}$ & $1.84^{+0.04}_{-0.04}$ & $2.29^{+0.66}_{-0.47}$ &\\
& $3.78\times10^{-17}$ & $4.27^{+0.86}_{-0.72}$ & $1.84^{+0.04}_{-0.03}$ & $2.29^{+0.66}_{-0.47}$ &\\
& $7.54\times10^{-17}$ & $4.37^{+0.88}_{-0.73}$ & $1.84^{+0.03}_{-0.03}$ & $2.24^{+0.58}_{-0.46}$ &\\
& $1.50\times10^{-16}$ & $4.37^{+0.88}_{-0.73}$ & $1.84^{+0.03}_{-0.03}$ & $2.19^{+0.57}_{-0.45}$ &\\
& $3.00\times10^{-16}$ & $4.37^{+0.88}_{-0.73}$ & $1.84^{+0.03}_{-0.03}$ & $2.24^{+0.58}_{-0.42}$ &\\
\hline
\multicolumn{5}{c}{H 2356-309} \\
\hline
& $6.00\times10^{-19}$ & $5.50^{+0.81}_{-0.71}$ & $1.41^{+0.06}_{-0.07}$ & $1.02^{+0.21}_{-0.15}$ &\\
& $1.20\times10^{-18}$ & $5.13^{+0.76}_{-0.66}$ & $1.44^{+0.05}_{-0.06}$ & $1.12^{+0.23}_{-0.19}$ &\\
& $2.39\times10^{-18}$ & $4.90^{+0.73}_{-0.63}$ & $1.46^{+0.05}_{-0.06}$ & $1.23^{+0.28}_{-0.23}$ &\\
& $4.76\times10^{-18}$ & $4.37^{+0.65}_{-0.56}$ & $1.50^{+0.05}_{-0.05}$ & $1.45^{+0.37}_{-0.27}$ &\\
& $9.50\times10^{-18}$ & $4.27^{+0.75}_{-0.55}$ & $1.54^{+0.04}_{-0.05}$ & $1.45^{+0.42}_{-0.30}$ &\\
& $1.89\times10^{-17}$ & $3.98^{+0.59}_{-0.51}$ & $1.57^{+0.04}_{-0.05}$ & $1.66^{+0.58}_{-0.37}$ &\\
& $3.78\times10^{-17}$ & $3.80^{+0.66}_{-0.49}$ & $1.59^{+0.04}_{-0.05}$ & $1.78^{+0.68}_{-0.43}$ &\\
& $7.54\times10^{-17}$ & $3.63^{+0.64}_{-0.54}$ & $1.61^{+0.04}_{-0.05}$ & $1.86^{+0.83}_{-0.48}$ &\\
& $1.50\times10^{-16}$ & $3.89^{+0.68}_{-0.58}$ & $1.61^{+0.04}_{-0.05}$ & $1.74^{+0.66}_{-0.42}$ &\\
& $3.00\times10^{-16}$ & $3.63^{+0.64}_{-0.54}$ & $1.62^{+0.04}_{-0.05}$ & $1.95^{+0.87}_{-0.50}$ &\\
\hline
\label{tab:best_fit_params_B_field}
\end{longtable}

\begin{multicols}{2}
\section{Spectra of sources that do not yield constraints on IGMF}
\label{sec:spectra}

\begin{figure*}
    \includegraphics[width=0.66\columnwidth]{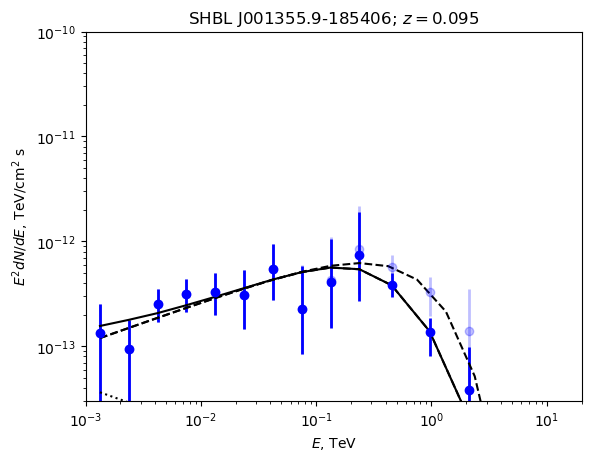}
    \includegraphics[width=0.66\columnwidth]{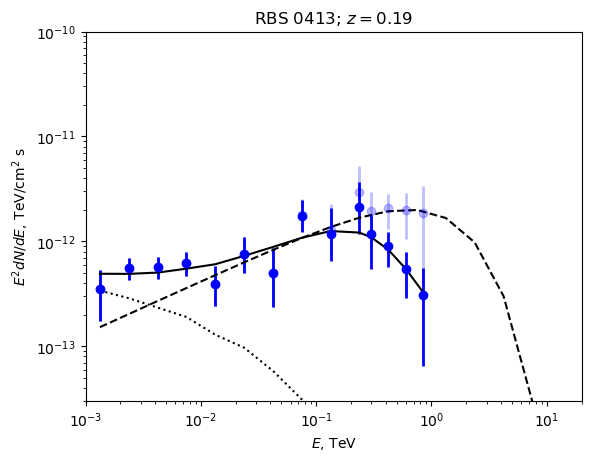}
    \includegraphics[width=0.66\columnwidth]{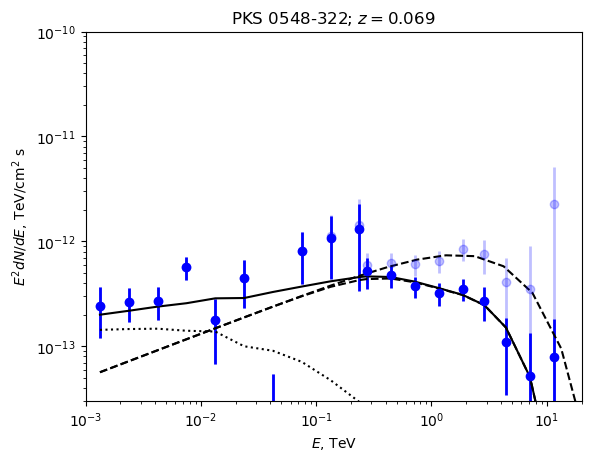}
    \includegraphics[width=0.66\columnwidth]{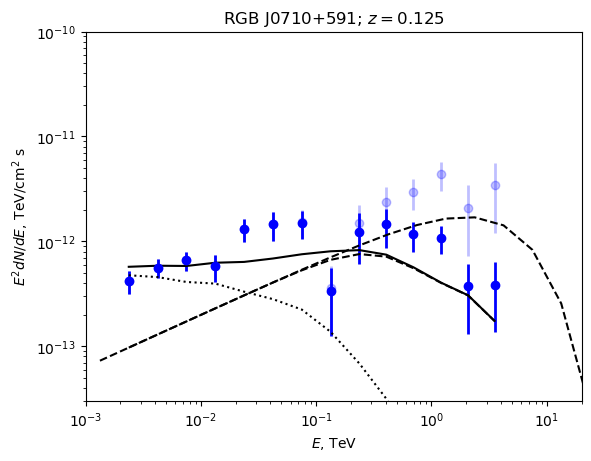}
    \includegraphics[width=0.66\columnwidth]{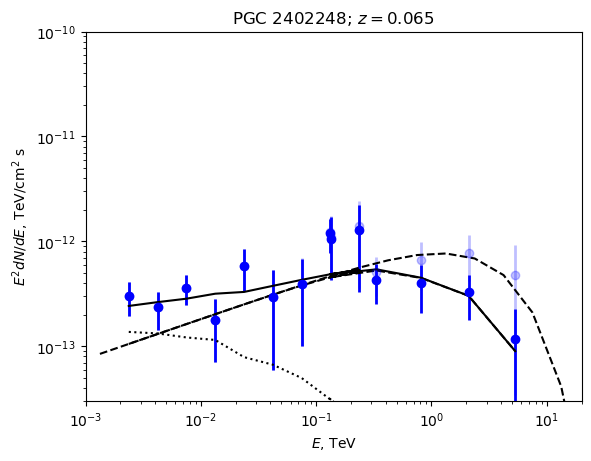}
    \includegraphics[width=0.66\columnwidth]{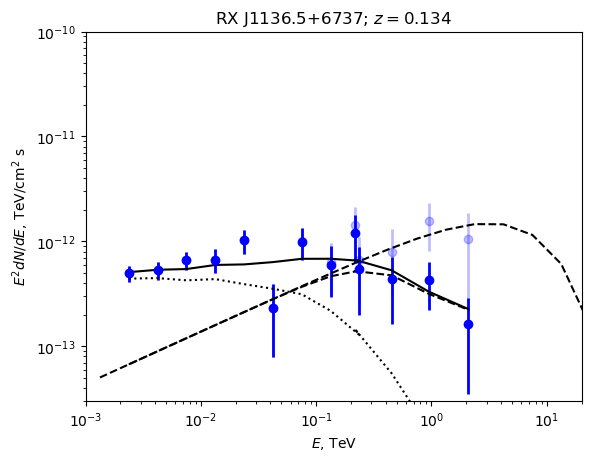}
    \includegraphics[width=0.66\columnwidth]{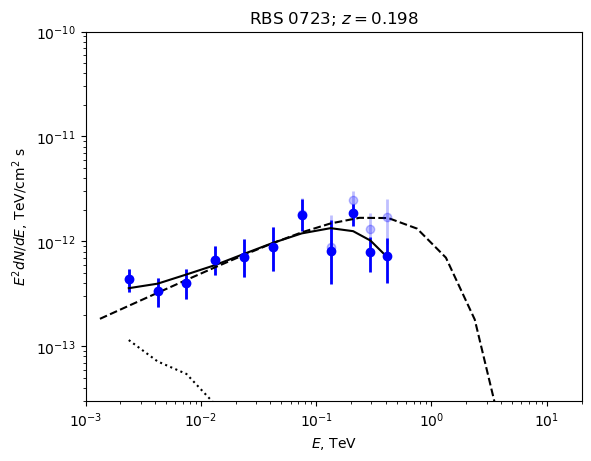}
    \includegraphics[width=0.66\columnwidth]{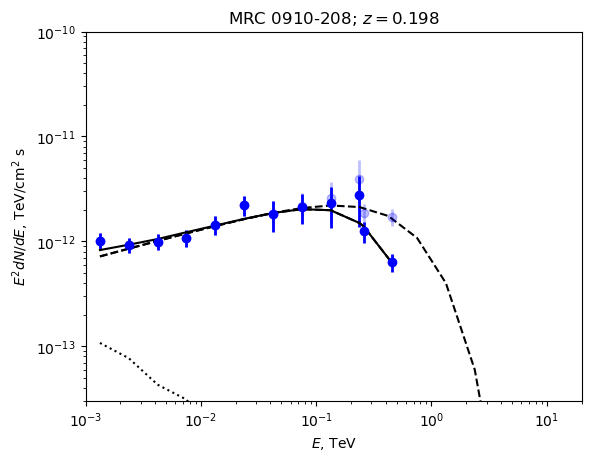}
    \includegraphics[width=0.66\columnwidth]{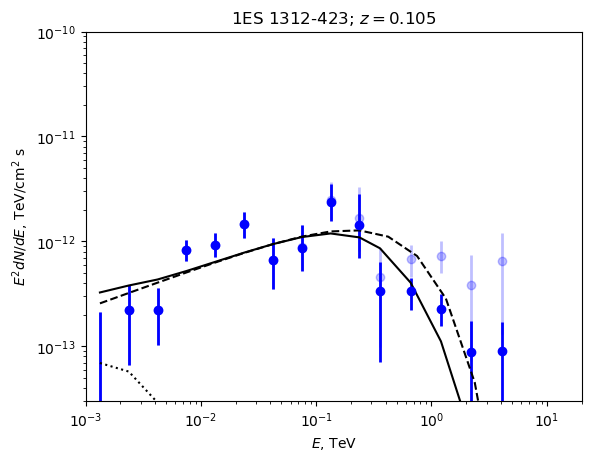}
    \includegraphics[width=0.66\columnwidth]{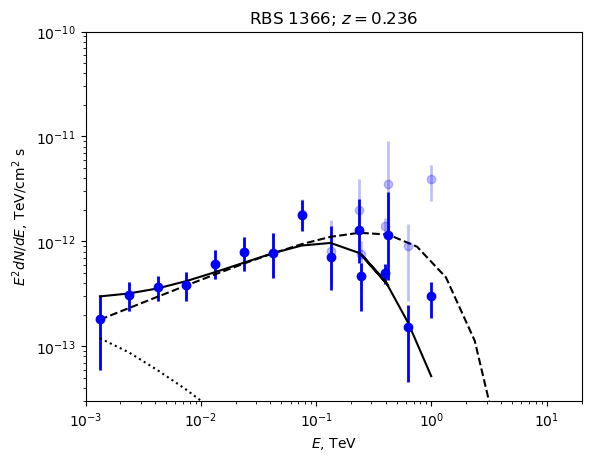}
    \includegraphics[width=0.66\columnwidth]{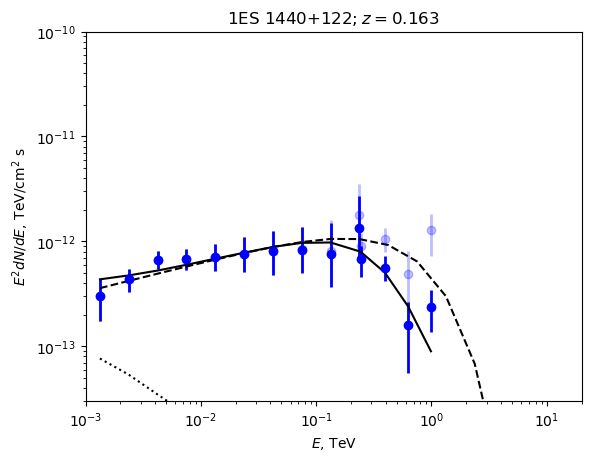}
    \includegraphics[width=0.66\columnwidth]{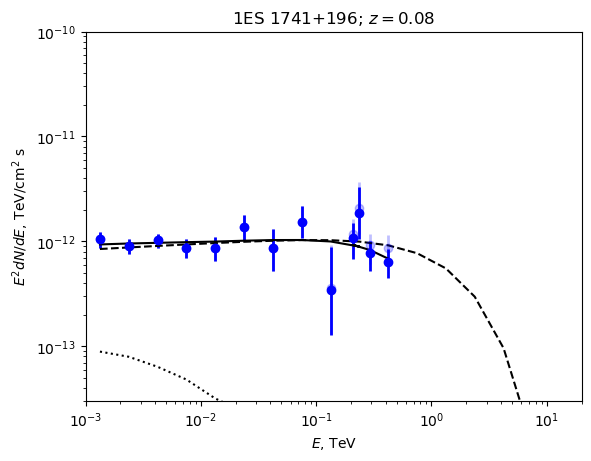}
    \includegraphics[width=0.66\columnwidth]{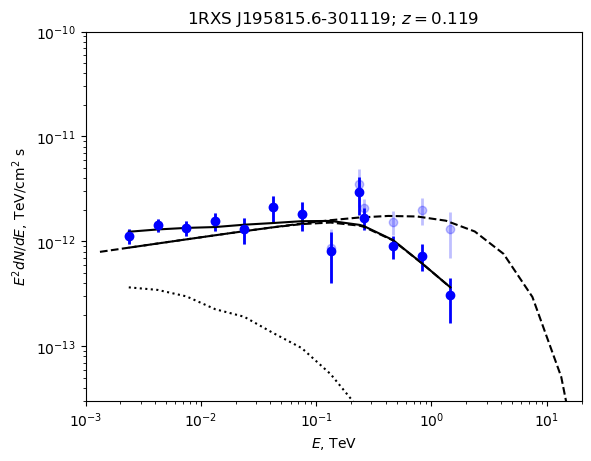}
    \includegraphics[width=0.66\columnwidth]{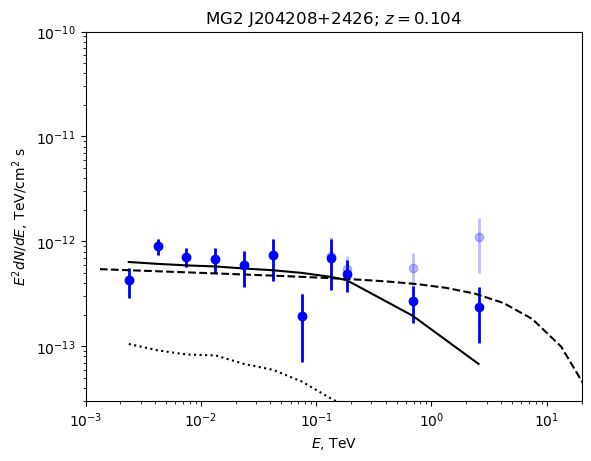}
   \caption{Same as in Fig. \ref{fig:spectra} but for other sources that do not yield constraints on IGMF. }
    \label{fig:candidates}
\end{figure*}

In this appendix we show spectral fits of the model with secondary flux for sources that do not currently yield constraints on IGMF. This can be either because the quality of the data is not sufficient for detection of the secondary flux or because the source spectrum has high-energy cut-off that suppresses the secondary flux. Fig. \ref{fig:candidates} shows the spectra of  such sources.

\end{multicols}

\end{appendix}

\end{document}